\documentclass[prd,preprintnumbers,amsmath,amssymb,floatfix,showpacs]{revtex4}

\usepackage{graphicx}
\usepackage{dcolumn}
\usepackage{bm}
\usepackage{amssymb}
\usepackage{epsfig}
\usepackage{color}

\newcommand{\ba}{\begin{eqnarray}}
\newcommand{\ea}{\end{eqnarray}}
\newcommand{\be}{\begin{equation}}
\newcommand{\ee}{\end{equation}}
\newcommand{\bdisplay}{\begin{displaymath}}
\newcommand{\edisplay}{\end{displaymath}}

\newcommand{\eq}[1]{Eq.\,(\ref{#1})}
\newcommand{\fig}[1]{Fig.\,\ref{#1}}

\newcommand{\delchi}{\Delta \chi^2_i}

\newcommand{\delchimax}{{\delchi}_{\rm max}}

\begin{document}

\title{Eikonal  and asymptotic fits to high energy data for $\sigma$, $\rho$, and $B$: An update with curvature corrections}

\author{Loyal Durand}
\email{ldurand@hep.wisc.edu}
\altaffiliation{Mailing address: 415 Pearl Ct., Aspen, CO 81611}
\affiliation{Department of Physics, University of Wisconsin, Madison, WI 53706}
\author{Phuoc Ha}
\email{pdha@towson.edu}
\affiliation{Department of Physics, Astronomy and Geosciences, Towson University, Towson, MD 21252}

\begin{abstract}

We update our eikonal fit and comprehensive asymptotic fits to high energy data on proton--proton and antiproton--proton scattering for $\sigma_{\rm tot}$, $\sigma_{\rm elas}$, $\sigma_{\rm inel}$, $\rho$, and $B$. The fits include the new TOTEM values of total proton-proton cross section, $\rho$, and $B$ at $W=\sqrt{s}$ = 13 TeV and the Telescope Array value of the total proton-proton cross section at $W=\sqrt{s}$ = 95 TeV,  data from the latest measurements of the inelastic cross sections at $W$= 8 TeV (by TOTEM and ATLAS) and 13 TeV (by CMS, ATLAS, and TOTEM). An important new feature of this work is the correction of the data to include the effects of curvature in $\ln{(d\sigma/dt)}$ on the values of $B$, $d\sigma/dt$ at $t=0$, and $\sigma_{\rm tot}$ obtained by extrapolation from the larger values of $t$ where the differential cross section is measured, The effects are significant. The stability of the fits is excellent and the new results agree well with the predictions of earlier fits. This work again confirms the evidence for the proton asymptotically becoming a black disk of gluons.

\end{abstract}


\date{\today}

\maketitle


\section{Introduction \label{sec:introduction}}

In a recent paper \cite{bdhh-eikonal}, we presented a detailed eikonal fit to the then-current data on  proton-proton and antiproton-proton scattering at center-of-mass energies $W=\sqrt{s}$ from 5 GeV to 57 TeV. The fit included data on the total and elastic scattering cross sections, the ratios $\rho$ of the real to the imaginary parts of the forward elastic scattering amplitudes, and the logarithmic slopes $B$ of the differential cross sections $d\sigma/dt$ at $t=0$.

In second paper \cite{bdhh-comprehensive}, we gave the results of a comprehensive fit to the data on the total, inelastic, and elastic scattering cross sections, $\rho$, and B for $pp$ and $\bar{p}p$ scattering  between 6 GeV and 57 TeV using parametrizations of those quantities which reflect the established $\ln^2{s}$ behavior of the cross sections at high energies \cite{blockcahn,blockrev}. The data were sufficient to show that $\sigma_{\rm elas}/\sigma_{\rm tot} \rightarrow 1/2$ at very high energies, and that  $8\pi B/\sigma_{\rm tot}\rightarrow 1$. These relations are exact for scattering from a black disk, and are satisfied in our eikonal model. The fact that they hold in experiment  provides strong evidence that the $pp$ and $\bar{p}p$ scattering amplitudes approach the black-disk limit asymptotically.

In the present paper, we update our eikonal fit and comprehensive asymptotic fits to high energy data on $pp$ and $\bar{p}p$ scattering for $\sigma$, $\rho$, and $B$, including the new TOTEM values of the $pp$ total cross section $\sigma_{\rm tot}=(110.6 \pm 2.3)$ mb, and $B=(20.36 \pm 0.19)$  GeV$^{-2}$ \cite{TOTEM2017}, and  $\rho=(0.1 \pm 0.01)$ \cite{TOTEM2017_2} at $W$ = 13 TeV;   $\sigma_{\rm tot}=(102.9 \pm 3.4)$ mb , $\rho=(0.12 \pm 0.03)$, and $B=(20.47 \pm 0.14)$  GeV$^{-2}$ at $W$ = 8 TeV \cite{TOTEM2016}; and
 the Telescope Array value of the total $pp$  cross section $\sigma_{\rm tot}=170 \ {}^ {+ 48}_{-44} ({\rm stat})^ {+ 19}_{-17} ({\rm syst.}) $ mb at $W$ = 95 TeV \cite{TA2017}. We also include data from the latest measurements of the inelastic cross sections at $W$=8 TeV, $\sigma_{\rm inel}=74.7 \pm 1.7 $ mb by TOTEM \cite{TOTEM2013_3} and $\sigma_{\rm inel}=71.73 \pm 0.15 ({\rm stat}) \pm 0.69 ({\rm syst}) $ mb by ATLAS \cite{ATLAS2016a}; and at $W$= 13 TeV, $\sigma_{\rm inel}=71.26 \pm 0.06 ({\rm stat}) \pm 0.47 ({\rm syst}) \pm 2.09 ({\rm lum}) \pm 2.72 ({\rm ext}) $ mb by CMS \cite{CMS2016}, $\sigma_{\rm inel}=78.1 \pm 0.6 ({\rm exp}) \pm 1.3 ({\rm lum}) \pm 2.6 ({\rm ext}) $ mb by ATLAS \cite{ATLAS2016b}, and $\sigma_{\rm inel}=79.5 \pm 1.8$ mb by TOTEM \cite{TOTEM2017}.

 In a new feature of this analysis, we include the corrections to the experimental values of $B$ and the total cross sections in the energy range 52--7000 GeV discussed in \cite{bdhh-curvature}. These result from the effects of curvature terms in $\ln(d\sigma/dt)$ which were not included in earlier experimental analyses, but affect the extrapolation of $\ln(d\sigma/dt)$ from the often fairly large values of $|t|$ or $q^2$ measured to $t=0$ to determine $B$ and $d\sigma(W,t)/dt|_{t=0}=(1+\rho^2)\sigma_{\rm tot}^2/16\pi$. These  terms were included by the TOTEM Collaboration in the recent analyses of their data at  8 TeV \cite{TOTEM2015,TOTEM2016} and 13 TeV \cite{TOTEM2017}.

 We find that our new fits are stable,  differing only slightly from the earlier results. The agreement with the data is impressive in both cases.  This work again confirms the evidence for the proton asymptotically becoming a black disk of gluons.

 We comment also on corrections to $\rho$, the problems we encounter in fitting the differential cross section at large momentum transfers and the highest energies, and on unitarity violations  and other difficulties encountered with some models which attempt to parametrize the differential cross sections directly.


\section{Data and corrections to simple exponential fits to $d\sigma/dq^2$ \label{sec:corrections}}


The data we use in our analysis consists of results on $\sigma_{\rm tot}$ for $W\geq 6$ GeV,  $\sigma_{\rm inel}$ for $W\geq 540$ GeV,  $\sigma_{\rm elas}$ for $W\geq 30$ GeV, and  $\rho$  for $W\geq 10$ GeV. The energy ranges for $\sigma_{\rm tot}$, $\sigma_{\rm inel}$, and $\rho$ are the same as those used in our previous work \cite{bdhh-comprehensive}, including
the new TOTEM values of total proton-proton cross section, $\rho$, and $B$ at $W$ = 13 TeV \cite{TOTEM2017}, the Telescope Array value of total proton-proton cross section at $W$ = 95 TeV \cite{TA2017} and the latest measurements of the inelastic cross sections at $W$= 8 TeV (by TOTEM \cite{TOTEM2013_3} and ATLAS \cite{ATLAS2016a}) and at 13 TeV (by CMS \cite{CMS2016}, ATLAS \cite{ATLAS2016b}, and TOTEM \cite{TOTEM2017}).

In addition to including new data, we have used the results of \cite{bdhh-curvature} to take into account approximately the effects on measurements of $B$ and $\sigma_{\rm tot}$ of the nonlinear ``curvature'' terms in the square of the invariant momentum transfer $q^2=|t|$ in the expansion
\be
\label{ln(dsig_dt)}
\ln(d\sigma/dq^2) = A-Bq^2+Cq^4-Dq^6+\cdots = A+Bt+Ct^2+Dt^3+\cdots.
\ee
These effects have been ignored in most experimental analyses, with $\ln(d\sigma/dq^2)$ assumed to vary strictly linearly with $q^2$ or $t$, with the experimental values $A_{\rm exp}$ and $B_{\rm exp}$ determined by least-squares fits to $d\sigma(W,q^2)/dq^2$ over a range of small $q^2$. The fits are then used in extrapolations of the nuclear part of the differential cross section to $q^2=t=0$ to determine $d\sigma(W,q^2)/dq^2|_{q^2=0}$ and, with the Coulomb-nuclear interference included, the ratio $\rho(W)$ of the real to the imaginary parts of the forward scattering amplitude and then  $\sigma_{\rm tot}$ through the relation
\be
\label{sig_tot}
\frac{d\sigma}{dq^2}(W,0) = \frac{1}{16\pi}(1+\rho^2)\sigma_{\rm tot}^2(W).
\ee

As we showed earlier \cite{bdhh-curvature}, the curvature-type effects from $C$ and $D$ are significant even for momentum transfers $q^2$ which are quite small, $q^2\lesssim 0.1$ GeV$^2$. These affect the local slope  $B(W,q^2)$ of $d\sigma/dq^2$ which increases as $q^2\rightarrow 0$. As a result,  the values $A_{\rm exp}$ and $B_{\rm exp}$ determined in fits over ranges of $q^2$ away from zero are too small. The shifts in $B$ are frequently well outside  the quoted experimental uncertainties. The shifts in $A$ and $\sigma_{\rm tot}$ are smaller, but still significant in some cases. Higher order terms in the expansion of $\ln{(d\sigma/dq^2)}$ are unimportant for $q^2\lesssim 0.10$ GeV$^2$ in realistic models, but must be taken into account for $q^2>0.1-0.15$ GeV$^2$.

We derived the general expressions for the curvature terms in \cite{bdhh-curvature}. When the real part of the elastic scattering amplitude is small, these can be expressed in terms of products of moments of the imaginary part of the amplitude, and are strongly constrained by the total cross section in both magnitude and energy dependence. In general, the lower-order terms are quite well-determined in eikonal fits to the scattering amplitude which reproduce $\sigma_{\rm tot}$ and $d\sigma/dq^2$ at small $q^2$.

The curvature effects were observed directly by TOTEM, first at  8 TeV \cite{TOTEM2015,TOTEM2016} and more recently at 13 TeV \cite{TOTEM2017}, in analyses which included $B$, $C$, and $D$, or $b_1,\,b_2,\,b_3$ in the TOTEM notation, in their fits to the observed differential cross sections. The values of $B$, $C$ and $D$ predicted by our eikonal model fitted to $pp$ and $\bar{p}p$ data from 5 GeV to 57 TeV \cite{bdhh-eikonal}, agreed well with the values of the parameters obtained in their analysis, even though $C$ and $D$ were not used in making our eikonal fit. For example, with the slightly modified eikonal fit described below, we find $B=20.26$ GeV$^{-2}$, $C=9.18$ GeV$^{-4}$, and $D=26.53$ GeV$^{-6}$ at 8 TeV compared to the values $20.47\pm 0.14$ GeV$^{-2}$, $8.8\pm 1.6$ GeV$^{-4}$, and $20\pm 6$ GeV$^{-6}$ found by TOTEM \cite{TOTEM2016}. We note that the range in $q^2$ used in the TOTEM analysis extends far enough, up to 0.19 GeV$^2$, that the next term in the series for $\ln(d\sigma/dq^2)$ is expected to enter and slightly decrease the effective value obtained for $D$.

Since most of the data on differential cross sections at lower energies are not precise or extensive enough to support direct experimental determinations of $C$ and $D$, we will adopt the procedure used in \cite{bdhh-curvature}, where we showed that the inclusion of curvature terms calculated using the earlier eikonal fit improved the fits to experiment at representative energies, and furthermore, that a simple semi-analytic expression gave corrections to $B$ and $\sigma_{\rm tot}$ consistent with the refitted values.

Our approximate expression for the correction to $B_{\rm exp}$ follows from the observation that, with $d\sigma/dq^2$ steadily increasing and curving upwards as $q^2$ decreases over the fitting interval  $q_{\rm max}^2\geq q^2\geq q_{\rm min}^2 $, the fitted value $B_{\rm exp}$ must match the local slope $B(W,q_0^2)$ of $\ln{(d\sigma/dq^2)}$  at a unique point $q_0^2$ inside the interval.  In terms of the series expansion of $\ln{(d\sigma/dq^2)}$ in \eq{ln(dsig_dt)},
\be
\label{B(q0^2)}
B(W,q_0^2) = B-2Cq_0^2+3Dq_0^4-\cdots.
\ee
Thus, at the matching point,

\be
\label{B_correction}
B  = B_{\rm exp}+2Cq_0^2-3Dq_0^4+\cdots
\ee
as stated in \cite{bdhh-curvature}, with the final terms giving the small correction to the experimental result needed to obtain $B$ at $q^2=0$.

The numerical fits to data in \cite{bdhh-curvature} showed that $q_0^2\approx 0.6q_{\rm min}^2+0.4q_{\rm max}^2$ for a selection of cross sections from 52.8 GeV to 8 TeV for $q^2$ intervals with $q_{\rm max}^2\lesssim 0.1$ GeV$^2$. The slight shift of $q_0^2$ from the central point in the interval reflects the steady increase in the differential cross section as $q^2$ decreases. As a check, we made a series of calculations in which we fitted ``data'' from the eikonal model to the exponential form $d\sigma/dq^2=exp(A_{\rm fit}-B_{\rm fit}q^2)$; we found that the approximation $B_{\rm fit}=B(W,q_0^2)$ with $q_0^2$ chosen as above is, in fact, quite accurate. We will therefore use the expression in \eq{B_correction} to adjust the data used in in the fits $B$ below. The corrections are generally a few percent, but range up to $\sim10\%$ in several cases where the $q^2$ ranges used in the experimental analyses were large.

The foregoing construction suggests that $A_{\rm fit}$ should be given approximately by $A(W,q_0^2)+B(W,q_0^2) q_0^2$, where we have made a linear extrapolation from the local amplitude at $q_0^2$  to $q^2=0$.  We found in the calculations above that this approximation is good and accurate enough for our purposes. Thus, using the expression
\be
\label{A(q0^2)}
A(W,q_0^2) = A-Bq_0^2+Cq_0^4-Dq_0^6-\cdots,
\ee
replacing $B_{\rm exp}$ by the expression in \eq{B(q0^2)}, and $A_{\rm fit}$ by the experimentally determined amplitude $A_{\rm exp}$, and solving for $A$, we find that
\be
\label{Acorr}
A \approx A_{\exp}+ Cq_0^4-2Dq_0^6+\cdots.
\ee

The corrections are quite small for $q_0^2$ small. Using the relation $d\sigma(W,q^2)/dq^2|_{q^2=0}=exp[A(W)]$ and \eq{sig_tot}, we find that the fractional change in $\sigma_{\rm tot}$ relative to the value given by $A_{\rm exp}$  is
\be
\label{deltaA}
\sigma_{\rm tot}/\sigma_{\rm tot,\,exp} \approx 1+( Cq_0^4-2Dq_0^6)/2.
\ee
This agrees with the results we obtained in \cite{bdhh-curvature} by directly refitting experimental data using curvature terms  $C$ and $D$ taken from the eikonal model.  The corrections are quite small, ranging from a fraction of a percent for most points to a maximum value of 2.5\%, and are within the experimental uncertainties. We do not have similar expressions for the corrections to $\sigma_{\rm elas}$ and $\sigma_{\rm inel}$, but would clearly expect those to be very small as well. We will ignore them.

Given these results and their stability over the energy range of interest, we have applied the corrections to all the $q^2$-dependent data used in the following analyses, using the ranges $q_{\rm min}^2\leq q^2\leq q_{\rm max}^2$ given by the experimenters, with the condition that $q_{\rm max}^2\leq 0.15$ GeV$^2$.

The potential corrections to $\rho$ are more complicated, as these involve the Coulomb-nuclear interference.  As emphasized recently by Pacetti, Srivastava, and Pancheri \cite{Pacetti}, the results for $\rho$ are sensitive to the very rapid decrease of the real part of the nuclear amplitude and the ratio ${\rm Re}f/{\rm Im}f$ away from the forward direction, with a change in sign well before the first diffraction minimum in $d\sigma/dq^2$. This decrease has been ignored in some analyses, or taken as much less rapid than is found in realistic models such as the model in \cite{Pacetti} or the eikonal model considered here. We intend to return to this problem in the future.


\section{Update of the Comprehensive fits \label{sec:comprehensive}}


\subsection{Fit without high-energy constraints \label{subset:fit_LEconstraints}}

We begin with an update on our global fits to the high-energy total, elastic, and inelastic $pp$ and $\bar{p}p$ scattering cross sections, and the ratios $\rho$ of the real to the imaginary parts of the forward elastic scattering amplitudes $f(s,t)$. As before, we use the parametrizations of $\sigma_{\rm tot}$, $\sigma_{\rm el}$, and $\rho$ for $pp$ and $\bar{p}p$ scattering introduced by  Block and Cahn \cite{blockcahn,blockrev},
\ba
\label{sigma0}
\sigma^{\rm 0}(\nu) &=& c_0+c_1\ln\left(\frac{\nu}{m}\right)+c_2\ln^2\left(\frac{\nu}{m}\right)+\beta\left(\frac{\nu}{m}\right)^{\mu-1} \, , \\
\label{sigma_tot}
\sigma_{\rm tot}^{\pm}(\nu) &=& \sigma^{\rm 0}(\nu) \pm\delta \left(\frac{\nu}{m}\right)^{\alpha-1}, \\
\label{sigma_el}
\sigma_{\rm elas}^{\pm}(\nu) &=& b_0+b_1\ln\left(\frac{\nu}{m}\right)+b_2\ln^2\left(\frac{\nu}{m}\right)+\beta_e\left(\frac{\nu}{m}\right)^{\mu-1} \pm\delta_e \left(\frac{\nu}{m}\right)^{\alpha-1} \, , \\
\label{rho}
\rho^{\pm} &=& \frac{1}{\sigma_{\rm tot}^{\pm}(\nu)}\left[\frac{\pi}{2}c_1+\pi c_2\ln\left(\frac{\nu}{m}\right) - \beta \cot\left(\frac{\pi\mu}{2}\right) \left(\frac{\nu}{m}\right)^{\mu-1} +\frac{4\pi}{\nu}f_+(0)
 \pm \delta\tan\left(\frac{\pi\alpha}{2}\right)\left(\frac{\nu}{m}\right)^{\alpha-1}\right],
\ea
where the upper and lower signs are for $pp$ and $\bar{p}p$ scattering, respectively. Here $\nu$ is the laboratory energy of the incident particle, with $2m\nu=s-2m^2=W^2-2m^2$ where $W$ is the center-of-mass energy and $m$ is the proton mass. The inelastic cross sections are given by the differences between the total and elastic cross sections, $\sigma_{\rm inel}^{\pm} = \sigma_{\rm tot}^{\pm} - \sigma_{\rm elas}^{\pm}$. They are therefore parametrized simply as the differences of the expressions in Eqs.\ (\ref{sigma_tot}) and (\ref{sigma_el}); no new parameters appear.

 The 13 parameters  in these expressions are not constrained at very high energies. We did use the two low-energy analyticity constraints on the cross sections found by  Block and Halzen \cite{blockhalzenfit,block_analytic} and Igi and Ishida \cite{Igi,Igi2} using finite-energy sum rules to fix the cross sections at 4 GeV and assure that the model connects smoothly to the low-energy region where the data are dense, namely
 \ba
 \label{constraint1}
 c_0+c_1\ln{(\nu_0/m)}+c_2 \ln^2{(\nu_0/m)}+\beta(\nu_0/m)^{\mu-1}=48.58\, {\rm  mb}, \\
 \label{constraint2}
 \delta(\nu_0/m)^{\alpha-1}=-8.405\, {\rm  mb},
 \ea
 where $\nu_0=7.59$ GeV corresponding to $W=4$ GeV.
  We also used the two new ratio constraints on the coefficients of the Regge-like terms discussed in  \cite{bdhh-comprehensive},
   \be
   \label{ReggeConstraints}
   \beta_e=0.302\,\beta,\quad \delta_e=0.203\,\delta,
   \ee
   so end up with 9 free parameters.

 In \cite{bdhh-eikonal} we included a similar expression for $B$ with 5 free parameters. We have since concluded that this was not really appropriate at present energies. When the real parts of the scattering amplitudes are small, as here, $B$ is given to very good approximation as the ratio of the second moment of the imaginary part of the scattering amplitude in impact-parameter space, asymptotically fourth-order in $\ln{\nu}$, to the total cross section. The ratio can only approach a second-order polynomial at very high energies while we used that form also down to 10 GeV. The five free parameters also provide too much flexibility in fitting the high-energy data.

In our initial calculations, we used the expressions in Eqs.\ (\ref{sigma0})-(\ref{rho}) to fit both the uncorrected and corrected data on the cross sections and $\rho$. As expected from the small size of the corrections to the cross sections, those results agreed within the uncertainties of the fits. The fit using the corrected data gave the asymptotic value $b_2/c_2=0.486\pm 0.062$ for the ratio $\sigma_{\rm el}/\sigma_{\rm tot}$, a value consistent within the uncertainty to the ratio 1/2 expected if the scattering amplitudes approach the so-called black-disk limit asymptotically. We regard this as strong evidence that the black-disk limit is reached at very high energies, with its effects already evident in the multi-TeV region.

Since a large fraction of the total $\chi^2$ in the fit arose from a few datum points, we used the sieve algorithm \cite{sieve,blockhalzenfit}  to better identify outlying points and remove them from the data set used in our final fit. The sieve procedure is based on a Lorentzian probability distribution adjusted to give results that agree very well with those from a Gaussian distribution in the absence of outliers, but which still eliminates the latter efficiently when they are present. The theory and details of the sieve procedure and various tests are given in  \cite{sieve}.

Using a cutoff Lorentzian $\chi^2$ of 6  \cite{sieve} to identify outlying points, the sieve eliminated 5 points from the corrected data set of 115 datum points (6 from the uncorrected data), including 1 total cross section, 1 elastic cross section, 1 inelastic cross section, and 2 values of $\rho$. This left 110 points overall with 9 parameters in the fit, thus 101 degrees of freedom to fit.
 The $\chi^2/{\rm d.o.f.}$ for the final Gaussian fit to the data with the outliers eliminated was 0.819, or, renormalized by the sieve factor ${\cal R}\approx 1.11$  \cite{sieve} to correct for the cutoff, ${\cal R}\chi^2/{\rm d.o.f.}=0.908$. This is an excellent fit, and gave a black-disk ratio $b_2/c_2=0.570\pm 0.108$, again consistent with the expected value 1/2. We therefore take the black-disk limit as established, and impose this as a further constraint in the analysis in the following section.

 We do not give separate lists of the parameters for this fit or the fit without the use of the sieve, or curves for the cross sections and $\rho$, as those parameters agree within statistics with the parameters obtained in the next section and given in Table\ \ref{table2:fit12p}, and the the curves for the cross sections and $\rho$ are nearly identical to the curves in Fig.\ \ref{fig:xsectionsBD}.


\subsection{Fit using the black disk constraint \label{subsec:black-disk}}

In our final fit, we used the general parametrization in Eqs.\  (\ref{sigma0})-(\ref{rho})  with both the low-energy constraints in Eqs. (\ref{constraint1}) and (\ref{constraint2}) and the high-energy black-disk constraint $b_2/c_2=1/2$ imposed, to fit the combined $pp$ and $\bar{p}p$ data over the same energy ranges as above. The sieve algorithm was again used to eliminate the same 4 outliers among 115 datum points. There are now only 8 parameters in the fit.

The result of the fit is excellent as seen in the last lines in Table \ref{table2:fit12p}, with a $\chi^2$ of 90.9 for 103 degrees of freedom for a raw $\chi^2$ per d.o.f. of 0.882 and a renormalized $\chi^2/{\rm d.o.f.}$ of 0.979. As would be expected, the parameters of the fit  have smaller uncertainties than in the previous fit using only the low-energy constraints, and change only within the previous uncertainties.

\begin{figure}[htbp]
\includegraphics{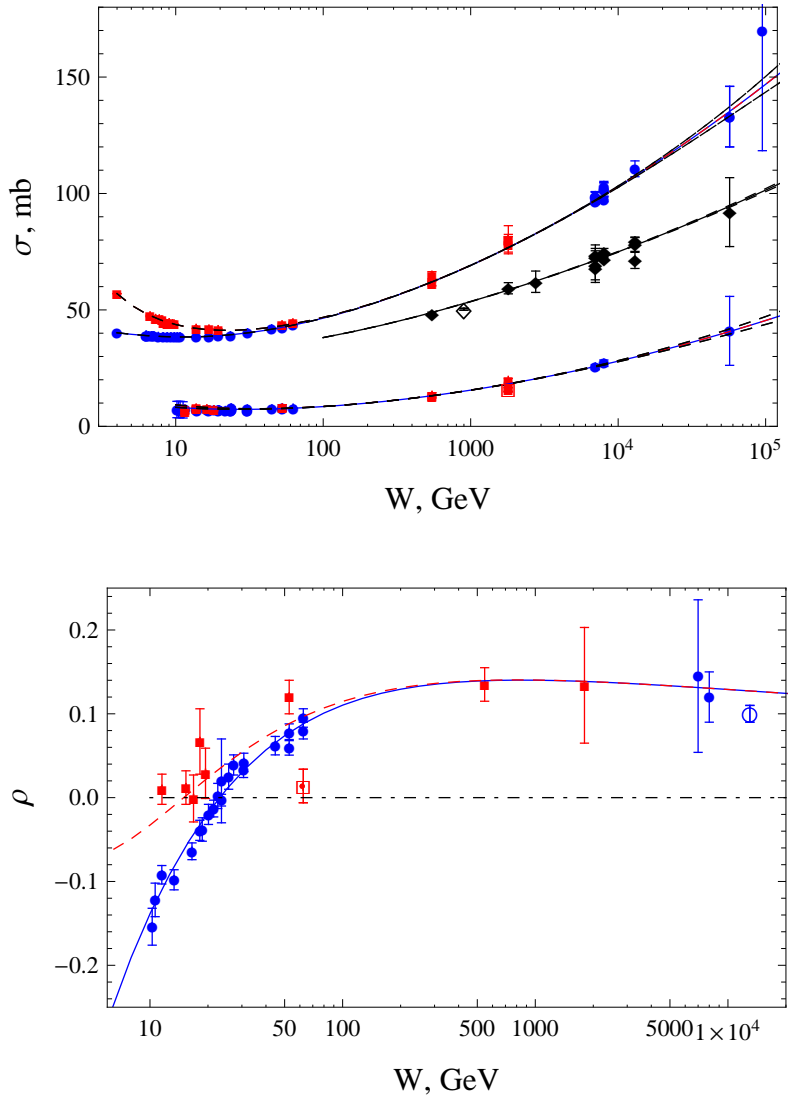}
\caption{Top figure: the fits, top to bottom, to the total, inelastic, and elastic scattering cross sections  using the high-energy black-disk constraint as well as the the low-energy analyticity constraints and the ratio constraints on the Regge-like contributions to the low-energy cross sections: $\sigma_{\rm tot}^{\bar{p}p}$ and  $\sigma_{\rm elas}^{\bar{p}p}$ (red squares and dashed red line); $\sigma_{\rm tot}^{pp}$  and $\sigma_{\rm elas}^{pp}$ (blue dots and solid blue line); $\sigma_{\rm inel}^{\bar{p}p}$ (black diamonds and line); $\sigma_{\rm inel}^{pp} $  (purple triangles and line).   The fit used only data  on $\sigma_{\rm tot}$ for $W\geq 6$ GeV, $\sigma_{\rm elas}$ for $W\geq 30$ GeV, and $\sigma_{\rm inel}$ for $W\geq 540$ GeV.  The curve for $\sigma_{\rm elas}$ includes data down to 10 GeV to show how the cross section is tied down at lower energies.  The statistical error bands determined by the error analysis are shown.  Bottom figure: the fit to $\rho$ for $\bar{p}p$ (red squares and dashed line) and $pp$ (blue dots and line) scattering. In both figures, outlying points identified in the sieve analysis and not used in the fit are shown with large open symbols surrounding the central points; the size of those symbols is not connected to the quoted errors.  }
\label{fig:xsectionsBD}
\end{figure}

 We give  combined plots of the total, inelastic, and elastic cross sections and $\rho$ at high energies in  Fig. \ref{fig:xsectionsBD}.  All the data used are shown, including the 2 cross section points and 2 values of $\rho$ which were dropped in the sieve analysis. We also show the statistical error bands for the fit; these show that the fit is very tightly constrained over the region of the data.
 The consistency with the fit without the high-energy constraints and the rather small uncertainty in $c_2 = 0.233 \pm 0.023$ mb indicate that the asymptotic cross sections are also well-determined.

%


\begin{table}[ht]                   
%
\def\arraystretch{1.15}            
\begin{center}				  
\begin{tabular}[b]{|l||c||}

\hline
{\rm Parameters} & $\delchimax=6$ \\
\hline
       $c_0$\ \ \   (mb) & $ 25.64 \pm 2.36 $ \\
      $c_1$\ \ \   (mb) & $0.158 \pm 0.647$ \\
      $c_2$\ \ \ \   (mb) & $0.233 \pm 0.023 $ \\
      $b_0$\ \ \   (mb) & $6.903 \pm 1.859 $ \\
      $b_1$\ \ \   (mb) & $-0.904 \pm 0.290$ \\
      $b_2$\ \ \    (mb) & $0.117 \pm  0.012 $ \\
      $\beta$\ \   (mb) & $44.84 \pm 4.00 $ \\
      $\beta_e$\ \  (mb) & $14.44 \pm 1.29 $ \\
      $f(0)$ (mb GeV) & $0.630 \pm 1.282$\\
      $\delta$\ \ (mb) & $-29.30 \pm 0.37 $\\
      $\delta_e$\ \  (mb) & $-5.95 \pm 0.08 $ \\
      $\alpha$ & $0.403 \pm 0.006$ \\
      $\mu $ & $0.651 \pm 0.032$ \\
\hline
\hline
	$\chi^2_{\rm min}$ & 90.9 \\
	${\cal R}\times\chi^2_{\rm min}$ & 100.8\\
	Degrees of freedom (d.o.f).&103\\
\hline
	${\cal R}\times\chi^2_{\rm min}$/d.o.f.&0.979\\
\hline
\end{tabular}
     \caption{\protect\small The results for our 8-parameter $\chi^2$ fit to the $\bar{p}p$ and $pp$ total, elastic, and inelastic cross sections, and $\rho$ values using the low-energy constraints, the black-disk constraint on the ratio $\sigma_{\rm el}/\sigma_{\rm tot}$, and  the cut $\delchimax=6$ in the sieve filtering of the data which eliminated 4 outlying points, 1 elastic and 1 inelastic cross section and 2 values of $\rho$. The renormalized $\chi^2_{\rm min}$/d.o.f.,  taking into account the effects of the Lorentzian $\delchimax$ cut is given in the row  labeled ${\cal R}\times\chi^2_{\rm min}$/d.o.f., with ${\cal R}(6)=1.110$ \cite{sieve}.   \label{table2:fit12p}}
\end{center}
\end{table}
\def\arraystretch{1}  

The crossing-even high energy inelastic cross section $\sigma^0_{\rm inel}(\nu)$, valid in the energy domain $\sqrt s \ge 100$ GeV where the odd Regge-like terms are very small and $\sigma_{\rm tot}^{pp}$ and $\sigma_{\rm tot }^{\bar{p}p}$ are essentially equal, is given by
\be
\sigma_{\rm inel}^0(\nu)= 18.76 + 1.062 \ln\left(\frac{\nu}{m}\right)
+ 0.1166 \ln^2\left(\frac{\nu}{m}\right)+30.40 \left(\frac{\nu}{m}\right)^{-0.3494} \ {\rm mb},
\label{finalinelastic}
\ee
the difference of the expressions for $\sigma_{\rm tot}$ and $\sigma_{\rm elas}$ with the coefficients in Table\ \ref{table2:fit12p}.

We note that the recent very precise TOTEM value of $\rho$, $\rho=0.1\pm0.01$ at 13 TeV, was rejected in the sieve analysis as seen in the lower panel in \fig{fig:xsectionsBD}, and again lies well off the nearly identical curves obtained with the complete data set with or without the black-disk constraint. The deviation of this point from the trend of the lower-energy data was interpreted by the TOTEM group \cite{TOTEM2017_2} as evidence for a crossing-odd ``odderon'' contribution to the scattering amplitude, as developed in detail by Martynov and Nicolescu \cite{Nicolescu}. This conclusion is not justified at present.

As shown by Pacetti, Srivastava, and Pancheri \cite{Pacetti},  the strong variations in  magnitude of the real part of the scattering amplitude and of the ratio $\rho(W,q^2)={\rm Re}f/{\rm Im}f$ over the Coulomb interference region significantly change the Coulomb-nuclear interference effects. Their reanalyses of the TOTEM data gives $\rho=  0.136$ and $\rho=0.134$ at 8 TeV and 13 TeV respectively, compared to the corresponding TOTEM values $\rho=0.12\pm0.03$ \cite{TOTEM2016} and $\rho=0.10\pm0.01$ \cite{TOTEM2017_2} shown in \fig{fig:xsectionsBD}. The modified values agree well with the predictions of the fit. These results are supported by the work of Kohara {\em et al.} \cite{kohara}, who obtained similar results in a less detailed analysis.

Pacetti {\em et al.} \cite{Pacetti} based their analysis on the Barger-Phillips type model \cite{BargerPhillips} of Fagundes {\em et al.} \cite{Fagundes} which fits the data on differential and total cross sections from ISR to LHC energies very accurately, and correctly predicted the differential cross section at 13 TeV. The model has the phase demanded by analyticity and crossing symmetry and the parametrized energy dependence built in. This leads, as noted, to a rapid decrease in ${\rm Re}f$ to a zero and change in sign at $q^2=|t|\approx 0.15$ GeV$^2$.   The eikonal model considered in the next section also has the correct phase relations for crossing and analyticity built in, and again gives a rapid decrease of $\rho(W,q^2)$ with increasing $q^2$, with a zero and change of sign at  $q^2=0.155$ GeV$^2$  at 13 TeV. The TOTEM group, in contrast, assumed constant or slowly decreasing values of $\rho(W,q^2)$ over the interference region in their analysis.


\section{Update on the Eikonal fit \label{sec:eikonal}}

\subsection{Fits to the cross sections, $B$, and $\rho$ \label{subsec:eikonal_fits}}

We have used the eikonal parametrizations of the $pp$ and $\bar{p}p$ scattering amplitudes given in \cite{bdhh-eikonal} to refit the combined data on $pp$ and $\bar{p}p$ total cross sections for $W\geq 6$ GeV and the elastic scattering cross sections, $\rho$, and $B$ for energies $W\geq 10$ GeV. The fit was constrained as described in \cite{blockrev} by fixing the values of the total cross sections at $W=4$ GeV to match the results obtained from the extensive low-energy data. This is the same general energy range with the same constraints as used in \cite{bdhh-eikonal}, but the data now include the new values of total proton-proton cross section, $\rho$, and $B$ at $W=$ = 13 TeV from TOTEM collaboration and the value of the total proton-proton cross section at $W=\sqrt{s}$ = 95 TeV from the Telescope Array collaboration. We also include data for the inelastic cross sections in the energy range 546 GeV - 57 TeV in the fit, including the new cross sections measured at 8 TeV \cite{ATLAS2016a,TOTEM2013_3} and 13 TeV \cite{ATLAS2016b,CMS2016,TOTEM2017} by the ATLAS, CMS, and TOTEM collaborations.

The values of $B$ and $\sigma_{\rm tot}$ used in the fit were corrected for curvature effects in the extrapolation to $q^2=0$ using the expressions in Eqs.\  (\ref{B_correction}) and (\ref{deltaA}), the values of the parameters $C$ and $D$ obtained in the earlier eikonal fit \cite{bdhh-eikonal}, and the $q^2$ ranges used in the respective experimental analyses. The corrections are most significant for $B$.

The fit was performed using the sieve algorithm \cite{sieve} to eliminate 11 outlying points among 199  total datum points.   Nine parameters were used in the fit leaving 179 degees of freedom, a total $\chi^2$ of 199.2, and a raw $\chi^2/{\rm d.o.f.} =  1.11$. This must be renormalized by the sieve factor ${\cal R}\approx 1.11$ to ${\cal R}\chi^2/{\rm d.o.f.}=1.22$ to account for the elimination of the outliers \cite{sieve}.  We note that {\em all} datum points including the outliers omitted in the final fit are shown in the figures comparing the fits with  data.

Our parametrization of the eikonal scattering amplitude is given in the appendix to \cite{bdhh-eikonal}. The values of the parameters found in the fit is given in Table \ref{tab:parameters}.


\begin{table}
\renewcommand{\arraystretch}{1.2}
 \caption{Summary of the parameters used in the fit to the $pp$ and $\bar{p}p$ scattering data in the eikonal model}
 \label{tab:parameters}
\begin{tabular}{| c c l |}
\hline
\hline
Fixed values && Fitted parameters \\
\hline
$m_0=0.6$ GeV && $C_0=7.386 \pm 0.07$ \\
$W_0=4$ GeV && $C_1=31.00 \pm 0.02$\\
$\mu_{gg}=0.705$ GeV && $C_2=-0.360 \pm 0.0004 $\\
$\mu_{qq}= 0.89 $ GeV && $C_3=-1.203 \pm 0.004 $ \\
$\mu_{odd}= 0.60$ GeV && $C_4=7.381\pm 0.013 $ \\
&& $C_5=-26.24 \pm 0.02$ \\
$\alpha_s=0.5$ && $\alpha_1=0.3196 \pm 0.0003 $ \\
 $\Sigma_{gg}=9\pi\alpha_s^2/m_0^2$ && $\alpha_2= 0.4640 \pm 0.0001 $ \\
$\  =19.635$ GeV$^{-2}$  && $\beta=0.1786\pm 0.0002$ \\
  \hline
\hline
 \end{tabular}
 \renewcommand{\arraystretch}{1}
 \end{table}

\begin{figure}[htbp]
\includegraphics{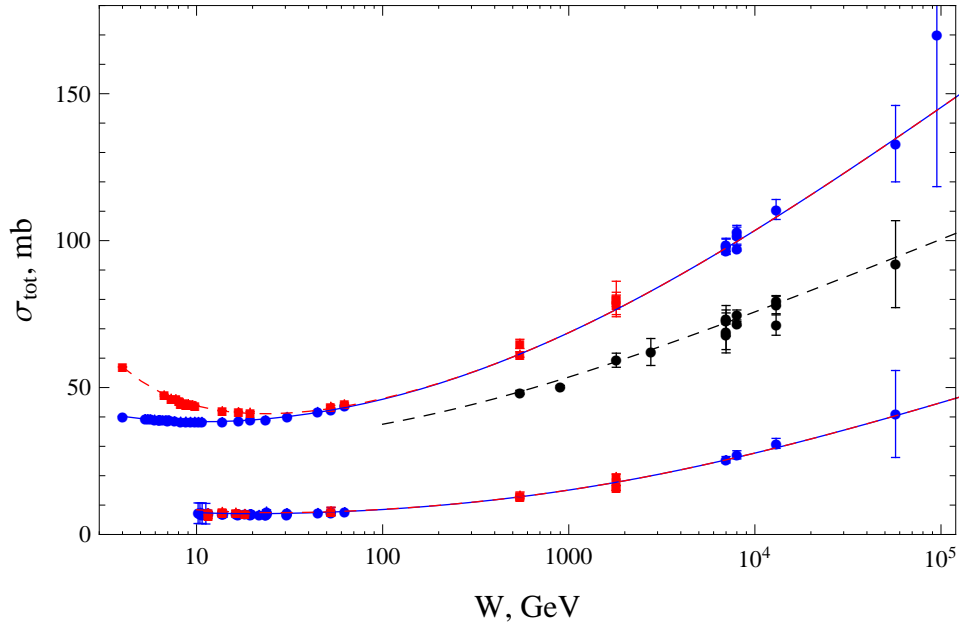}
\caption{Eikonal  fits to $\sigma_{\rm tot,pp}$ (blue dots and solid line)  and $\sigma_{\rm tot,\bar{p}p}$ (red squares and dashed line).  Only data above 5 GeV were used in the final fit, with the cross sections constrained to fit compilations of low-energy data at 4 GeV \cite{blockrev}. }
\label{fig:crosssectionfits}
\end{figure}

The results for the fits to the total, inelastic, and elastic scattering cross sections are shown in \fig{fig:crosssectionfits}. The fits are excellent, and very close  to those obtained in Sec.\ \ref{subsec:black-disk} using the Block-Cahn parametrization \cite{blockcahn,blockrev} of their expected high-energy behavior with the black-disk constraint. However, the eikonal model is more informative in that it allows the calculation of more quantities of experimental interest including differential cross sections and the curvature parameters discussed earlier.

\begin{figure}[htbp]
\includegraphics{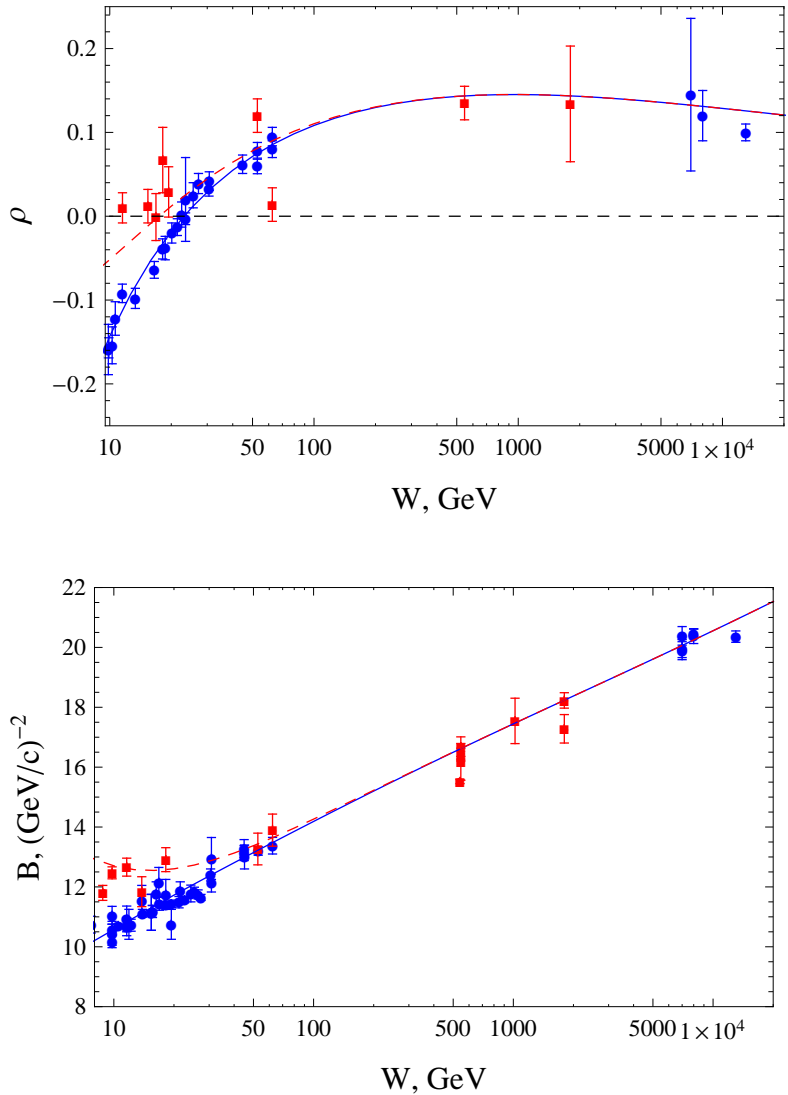}
 \caption{Top panel: eikonal  fits to the ratios $\rho$ of the real to the imaginary parts of the forward scattering amplitudes for $pp$ (blue dots and solid line) and $\bar{p}p$ (red squares and dashed line) scattering. The horizontal dashed line is at $\rho=0$. Bottom panel: fits to the logarithmic slopes of the elastic differential scattering cross sections $d\sigma/dt$ for $pp$ (blue dots and solid line) and $\bar{p}p$ (red squares and dashed line) scattering. }
   \label{fig:rhoBfits}
\end{figure}

The fits to the logarithmic slopes $B$ of the forward differential elastic scattering cross sections $d\sigma/dt$, are shown in \fig{fig:rhoBfits}.  The data for $B$ include the TOTEM results \cite{TOTEM2013,TOTEM2013_2}  at  $W=8$ TeV where curvature corrections were included in the experimental analysis. The results of that analysis gave values for $B$, $C$, and $D$ in agreement with the predictions of the eikonal model, as already noted. The parameters of the TOTEM analysis at 13 TeV were unfortunately not included in \cite{TOTEM2017_2}, and the range of $q^2$ used in their fit extends beyond that for which the result in \eq{B_correction} is reliable as determined in \cite{bdhh-curvature}. The next term in the series in \eq{ln(dsig_dt)} is expected to be significant at the larger values of $q^2$ in the range used, and act to decrease the effective value of $D$.

The fit to $\rho$ is also shown in \fig{fig:rhoBfits}.  The highest energy data for $\rho$ are from the TOTEM Collaboration at 13 TeV.  The TOTEM values of $\rho$ obtained at 8 TeV and, especially, at 13 TeV appear to lie well below the trend of the lower energy data. The value at 13 TeV, quoted with very low uncertainty, is excluded in the sieve analysis, though its inclusion in our fit makes very little difference in the results because of the preponderance of other data and the constraints imposed by the cross sections.

As discussed at the end of  Sec.\ \ref{subsec:black-disk}, the low values of $\rho$ have been shown by Pacetti, Pancheri, and Srivastava \cite{Pacetti} to result from the neglect in the TOTEM analysis \cite{TOTEM2017_2} of the strong $q^2$ dependence of the real part of the scattering amplitude, with ${\rm Re}f$ having a diffraction zero and changing sign at the very low value $q^2\approx 0.15$ GeV$^2$, within the $q^2$ range used in the analysis. The corrected values of $\rho$ obtained in \cite{Pacetti} are 0.136 at 8 TeV and 0.134 at 13 TeV, with some uncertainty. The eikonal model predicts a changes in sign of ${\rm Re}f$ at $q^2=0.17$ and 0.155 GeV$^2$, and values of $\rho$ of 0.131 and 0.126 at 8 and 13 TeV, quite consistent with the analysis of Pacetti {\em et al.}. We conclude that there is no reason to be concerned at this point about the the exclusion of the 13 TeV point in our fit, and no need to include an odderon contribution in the scattering amplitude as proposed in \cite{TOTEM2017_2} and \cite{Nicolescu}.

%

%


\begin{figure}[htbp]
\includegraphics{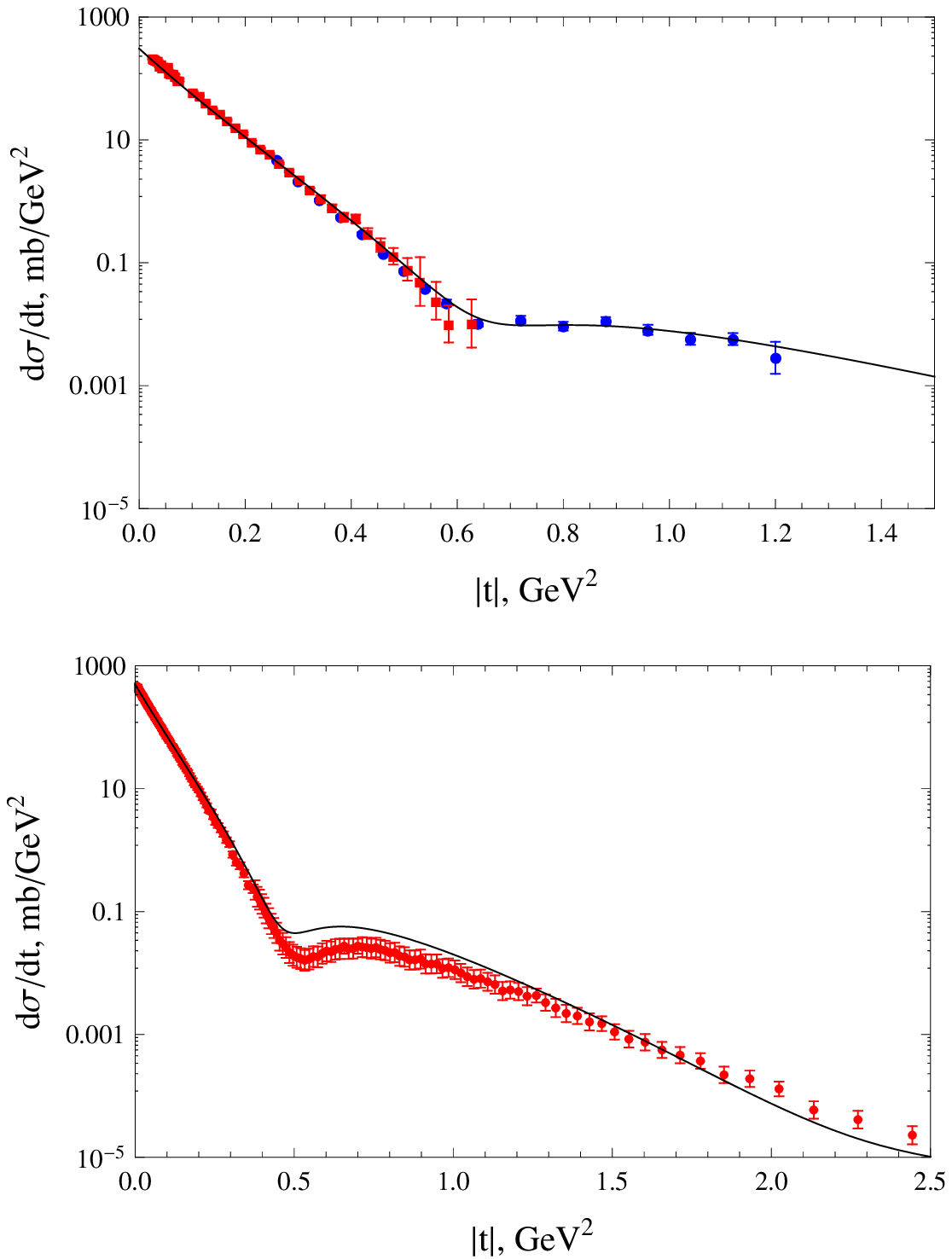}
 \caption{Top: differential cross section $d\sigma/dt$ calculated in the eikonal model compared to that from the E710 experiment \cite{E710_B,E710_B2} at $W=1800$ GeV. Bottom $d\sigma/dt$ from the eikonal model and the TOTEM experiment \cite{TOTEM2011} at $W=7000$ GeV.}
 \label{fig:dsigdt}
 \end{figure}
 %

The differential cross sections  were not used directly in our overall fits. The latter use only the information encoded in $\sigma_{\rm tot}$, $B$, and $\rho$ defined at small $q^2$. As is evident in the figures, the fits also compromise among datum points of comparable stated accuracy  that may disagree by amounts larger than the quoted uncertainties, so we do not expect to match individual differential cross sections exactly even for $q^2$ small.

The measured and predicted cross sections $d\sigma/dq^2$ are shown in \fig{fig:dsigdt}  at $W=1.8$ and 7 TeV. The results at 546 GeV and 62 GeV are comparable to those at 1.8 TeV.  Our descriptions of the cross sections at small $q^2$ are good at all energies, corresponding to our fits to the $B$ and $\rho$ parameters and total cross sections.  The predicted locations of the diffraction minimum is reproduced properly at 1.8 TeV and below, but is shifted slightly toward smaller $q^2$ relative to experiment at 7 TeV.  This pattern persists at the higher energies, with the predicted diffraction minimum shifted slightly toward smaller $q^2$ than observed, and the following peak somewhat too high.

The lack of precision near the diffraction minimum is not surprising. The minimum results from the vanishing of the imaginary part of the scattering amplitude caused by cancellations  between contributions from large and small impact parameters in the oscillating impact-parameter integral for that  quantity. This was discussed in \cite{bdhh-eikonal}.

The changes needed to correct this problem are  small: calculation shows that an addition to the imaginary part of the amplitude near $q^2=0.55$ GeV$^2$ of $\sim 0.7 \% $ of its value at $q^2=0$ would shift the minimum at 7 TeV to the proper location and reduce the height of the following maximum. A detailed fit would require finer modeling of the shape of the eikonal function than we have attempted so far, with an emphasis on the cancellations involving the terms which are  the dominant at the higher energies.

It is interesting in this connection to note that a different set of problems is encountered with models which attempt to fit the differential cross section directly using analytic expressions. An example is given by the Regge-type model of Donnachie and Landshoff \cite{Donnachie} which seemingly fits  $d\sigma/dt$ and $\sigma_{\rm tot}$ very well from $\sim$20 to 8000 GeV. However, a close examination shows that the values of $B$ derived from the model do not vary properly with energy over the lower part of this energy range, and the impact-parameter amplitudes derived from the the model amplitude by inverse Fourier-Bessel transformation are inconsistent with the eikonal form at high energies, hence violate unitary, a possibility of which those authors were aware.


\section{Conclusions \label{sec:conclusions}}

We have updated our eikonal fit and comprehensive fits to high energy data on proton--proton and antiproton--proton forward scattering for $\sigma$, $\rho$, and $B$, including the Telescope Array value of total proton-proton cross section at $W$ = 95 TeV and the latest measurements of the inelastic cross sections at $W$= 8 TeV (by TOTEM and ATLAS)  and 13 TeV (by ATLAS, CMS, and TOTEM). A new feature of the analysis is our inclusion of corrections to the reported values of $B$ and $\sigma_{\rm tot}$  associated with the effects of curvature in $\ln(d\sigma/dq^2)$ on the extrapolation from the measured range of $q^2$ to $q^2=0$ \cite{bdhh-curvature}. We give semi-analytic expressions for the corrections, and have implemented them using the earlier eikonal model \cite{bdhh-eikonal}; the results are not changed significantly in our updated eikonal fit.

We find that the fits agree well numerically and graphically with  our earlier works. The stability of the fits is not unexpected given the general agreement of the new data with our original predictions. The comprehensive fit using the Block-Cahn asymptotic parametrization of the cross sections and $\rho$ again gives an asymptotic ratio of $\sigma_{\rm elas}$ to $\sigma_{\rm tot}$ consistent within rather small uncertainties with 1/2, strongly indicating that the scattering approaches the black disk limit at very high energies. Earlier results on the ``edge'' of the scattering amplitude and black disk limit in \cite{bdhh-eikonal} are unchanged in the updated eikonal fit.  We find, however, that there are still problems in fitting $d\sigma/dq^2$ near the diffraction minimum where the scattering amplitude is very sensitive to small changes in the cancellations in the impact-parameter integrals which lead to the minimum.  Some small changes in the shape of the eikonal function are clearly needed.


\begin{acknowledgments}

L.D.\  would  like to thank the Aspen Center for Physics for its hospitality and for its partial support of this work under NSF Grant No. 1066293.    P.H.\ would like to thank Towson University Fisher College of Science and Mathematics for support.

\end{acknowledgments}


\bibliography{small_x_references_Oct_2018}

\end{document}